\documentclass[submission,copyright,creativecommons]{eptcs}
\providecommand{\event}{PxTP 2019} 
\usepackage{breakurl}             
\usepackage{underscore}           

\usepackage{algorithm}
\usepackage{algpseudocode}
\usepackage{qtree}

\usepackage{listings}
\usepackage{graphicx}
\usepackage[table]{xcolor}
\usepackage{tabularx}
\usepackage{booktabs}
\usepackage{amsfonts}
\usepackage{amssymb}
\usepackage{mathtools}
\usepackage{bussproofs}
\usepackage{caption}

\usepackage{hyperref}

\newtheorem{definition}{Definition}
\usepackage{lmodern}
\newcolumntype{C}[1]{>{\centering\arraybackslash}p{#1}}

\title{CLS-SMT: Bringing Together Combinatory Logic Synthesis and Satisfiability Modulo Theories}

\author{Fadil Kallat
\qquad\qquad
Tristan Sch\"afer
\qquad\qquad
Anna Vasileva
\institute{Technical University of Dortmund,\\
Dortmund, Germany}
\email{\{fadil.kallat, tristan.schaefer, anna.vasileva\}@tu-dortmund.de}}

\def\titlerunning{Bringing Together (CL)S and SMT}
\def\authorrunning{F. Kallat et al.}

\begin{document}
\maketitle

\begin{abstract}
We introduce an approach that aims to combine the usage of satisfiability modulo theories (SMT) solvers with the Combinatory Logic Synthesizer (CL)S framework.
(CL)S is a tool for the automatic composition of software components from a user-specified repository.
The framework yields a tree grammar that contains all composed terms that comply with a target type.
Type specifications for (CL)S are based on combinatory logic with intersection types.
Our approach translates the tree grammar into SMT functions, which allows the consideration of additional domain-specific constraints.
We demonstrate the usefulness of our approach in several experiments. 

\end{abstract}

\section{Introduction}
In component-based software synthesis, programs are not build from scratch but composed from a repository of typed combinators.
Combinators help to reduce the search space so that the inherent complexity of software synthesis problems can be handled.
Moreover, additional domain-specific knowledge is contained in the semantic type layer of a repository.
The underlying type system is well suited to express feature vectors of programs and software components.
A user-specified repository $\Gamma$ includes typed combinators that represent software components ($c\ :\ \sigma$) where $c$ is the component name and $\sigma$ is an intersection type \cite{DBLP:conf/isola/BessaiDDMR16,Bessai.2014}.

The Combinatory Logic Synthesizer (CL)S is a synthesis framework based on a type inhabitation algorithm for combinatory logic with intersection types \cite{JR13,DBLP:conf/isola/BessaiDDMR16}.
The algorithm searches for terms that are formed from the combinators and have a given target type $\tau$.
(CL)S is intended to be used for the automatic composition of software \cite{Mixins,FeatureGrammars,DBLP:conf/isola/BessaiDDMR16,SPLC15,DBLP:conf/isola/HeinemanBDR16}.
Besides the synthesis from software components, the (CL)S framework allows the synthesis of data structures, for instance of BPMN 2.0 processes \cite{DBLP:conf/isola/BessaiDDMR16} or planning processes \cite{winkels2018automatic}.

Obviously, the expression of domain-specific knowledge is limited by the underlying type system.
Intersection types do not explicitly take the logical connectives conjunction, disjunction and negation into consideration.
Moreover, the input-output behaviour of the resulting program cannot be expressed by types.
The combinatory approach allows to specify local typing information of a combinator but lacks expressivity regarding the global structure of result terms.
For instance, it is not possible to state that a combinator $c_0$ must contain combinator $c_1$ anywhere in the subtree of its arguments.
In some situations, not all well-formed terms might be considered to be reasonable results.
Different terms might also show identical execution results and runtime behaviour.

Software synthesis is an established research topic that offers a broad range of specification formalisms such as examples \cite{Zdancewic16,GHS12,SinghGulwani16,Udupa13}, types \cite{Zdancewic16,GKKP13,PKS16} or first-order-logic \cite{Reynolds17,TGD15}.
For this paper, we followed the intuition that the joint usage of (complementary) formalisms can yield a synthesis approach that combines the respective strengths of the underlying techniques.
Precisely, we identified SMT to be well working with combinatory logic.
There are different possible scenarios to incorporate these techniques.
For example, SMT could generate parts of combinators or parametrize synthesized programs.
In this paper, we show how to use SMT to filter a complete enumeration of inhabitants.
We implemented our approach in a tool called CLS-SMT.

The combinatory logic synthesis yields a tree grammar that describes the set of valid inhabitants.
We use this grammar to automatically construct a set of adequate SMT formulas.
By solving these formulas, we receive a tree model that represents a word of the grammar.
The (possibly infinite) set of inhabitants is further narrowed by introducing domain-specific structural constraints on terms.
That way, we can regulate the selection of result programs while avoiding trivial solutions.

The paper is organized as follows: In Section 2 we briefly introduce the composition synthesis framework (CL)S, its underlying theoretical background and the formalism of tree grammars.
Section 3 includes a presentation of CLS-SMT and the details about the translation of tree grammars into SMT formulas.
In Section 4 we evaluate our approach considering an example for sort programs and a labyrinth example.
Section 5 includes an overview of related work.
Finally, the conclusion gives a brief summary.

\section{Combinatory Logic Synthesizer (CL)S}
The developing tool Combinatory Logic Synthesizer (CL)S provides an implementation of a type inhabitation algorithm for combinatory logic with intersection types that is fully integrated into the Scala programming language.
The framework is publicly available \cite{combinatorsorg}.

The automatic software synthesis is performed by answering the type inhabitation question:
$ \Gamma \vdash ? : \tau $.
The problem of inhabitation asks for all well-typed applicative terms that can be formed from typed combinators in a user-specified set $\Gamma$ and have a given type $\tau$.
Applicative terms are defined as: $$M,N\ ::= c\ |\ (MN)$$
A term is constructed by using named component or combinator $c$ and application of $M$ to $N$, ($MN$).
If there exists a combinatory expression $M$ such that $ \Gamma \vdash M : \tau$ then $M$ is called inhabitant of $\tau$.
The type expressions that represent the specifications of term $M$ are denoted $\sigma$, $\tau$ and are defined as follows:
$$\sigma, \tau ::= a\ |\ \alpha \  |\ \sigma \to \tau\ | \ \sigma\ \cap\ \tau$$

Type constants ($a$) can be native or semantic types.
Type variables ($\alpha$) are substituted with type constants and facilitate generic components.
Furthermore, types can be constructed from function types ($\sigma \to \tau$) or intersections ($\sigma \cap \tau$).

There are four rules that control the type inhabitation process.
According to these rules, types are assigned to combinatory terms \cite{DBLP:conf/csl/DudderMRU12}.
The first rule (\textsf{var}) allows the usage of any combinator $c$ from the typed repository $\Gamma$ that has type $\tau$ using substitutions.
It is defined as follows:
\begin{prooftree}
 		\AxiomC{}
        \RightLabel{$(\textsf{var})$}
        \UnaryInfC{$\Gamma,c : \tau \vdash c : \mathcal{S}(\tau)$}
\end{prooftree}
Furthermore, it allows to assume that this combinator $c$ has type $S(\tau)$, where $S$ is a well-formed substitution on $\Gamma (c)$ mapping type variables to simple types.
The inhabitation problem in general is undecidable.
A restriction on variable substitution is needed to ensure decidability  \cite{DBLP:conf/csl/DudderMRU12}.

The following rule, arrow elimination $(\to\textsf{E})$, allows the application of combinators with function types to appropriately typed arguments to form terms.
 \begin{prooftree}
        \AxiomC{$\Gamma \vdash M : \sigma \to \tau$}
			\AxiomC{$\Gamma \vdash N : \sigma$}
        \RightLabel{$(\to\textsf{E})$}
        \BinaryInfC{$\Gamma \vdash MN : \tau$}
\end{prooftree}
The intersection introduction rule $(\cap \textsf{I})$, shown below, allows to type a term $M$ with two types, if there are proofs that $M$ has type $\sigma$ and type $\tau$.

\begin{prooftree}
        \AxiomC{$\Gamma \vdash M : \sigma$}
			\AxiomC{$\Gamma \vdash M : \tau$}
        \RightLabel{$(\cap \textsf{I})$}
        \BinaryInfC{$\Gamma \vdash M : \sigma \cap \tau$}
\end{prooftree}
The fourth rule $(\leq)$ deals with subtyping.

\begin{prooftree}
        \AxiomC{$\Gamma \vdash M : \sigma$}
			\AxiomC{$\sigma \leq \tau$}
        \RightLabel{$(\leq)$}
        \BinaryInfC{$\Gamma \vdash M : \tau$}
\end{prooftree}
The subtyping rules are based on the Barendregt-Coppo-Dezani-Ciancaglini (BCD) \cite{bcd} subtyping relation.
These include for example:
\begin{prooftree}
        \AxiomC{$A_2 \leq A_1$}
			\AxiomC{$B_1 \leq B_2$}
        \BinaryInfC{$A_1 \to B_1 \leq A_2 \to B_2$}
\end{prooftree}
to allow co- and contra-variant subtyping of functions and

\begin{center}
        \AxiomC{}
        \UnaryInfC{$A \cap B \leq A$}
  		\DisplayProof\hspace{2.5em}
        \AxiomC{}
        \UnaryInfC{$A \cap B \leq B$}
        \DisplayProof
\end{center}
to have intersection as the least upper bound.
The BCD system is also extended with type constructors, which was proposed in \cite{Laurent:18,Bessai:19}.


\subsection{Tree Grammar}

The (CL)S framework recursively computes all possible solutions in form of tree grammars \cite{Bessai.2018}.
We consider the generalized case of $normalized\ regular\ tree\ grammars$, which are well-known from literature \cite{tata}.

\begin{definition}{(Tree Grammars, Tree Grammar Languages)}

A tree grammar G is a 4-tuple $(S, \mathcal{N}, \mathcal{F}, R)$ with
\begin{itemize}
\item a start symbol $S \in \mathcal{N}$
\item a set $\mathcal{N}$ of \textit{nonterminals},
\item a set $\mathcal{F}$ of \textit{terminal symbols},
\item a set $R$ of \textit{productions rules} of form $\alpha_1 \mapsto \{c_1(\beta_1, \beta_2, \dots \beta_n),\ c_2(\gamma_1, \gamma_2, \dots \gamma_m)\}$,
 where $n, m \geq 0$, $\alpha_1, \beta_1, \beta_2, \dots, \beta_n, \gamma_1, \gamma_2,\dots, \gamma_m \in \mathcal{N}$ are nonterminal and $c_1, c_2 \in \mathcal{F}$ are terminal symbols.

We consider tree grammars without restriction on the arity of the terminal symbols, e.g. we can have $\alpha_1 \mapsto c_1(\beta_1, \beta_2)$ and $\alpha_2 \mapsto c_1(\beta_1)$ with $\alpha_2 \in \mathcal{N}$.

\end{itemize}
For a given tree grammar $G = (S, \mathcal{N}, \mathcal{F}, R)$ and nonterminal $\alpha \in \mathcal{N}$, $\mathcal{L}_{\alpha}(G)$ is the least set closed under the rule
\begin{align*}
    &\text{ if } \alpha \mapsto c(\beta_1, \beta_2, \dots, \beta_n) \in R \text{ and for all } 1 \leq k \leq n: t_k \in \mathcal{L}_{\beta_k}(G) \text{ then }\\
    &c(t_1, t_2, \dots, t_n) \in \mathcal{L}_{\alpha}(G)
\end{align*}
We define $\mathcal{L}(G) = \mathcal{L}_{S}(G)$ to be the language of grammar G.
\end{definition}

For request $\Gamma \vdash ?: \tau$, (CL)S constructs a tree grammar $G = (\tau,\mathcal{N}, \mathcal{F}, R)$ where $\tau \in \mathcal{N}$.
The right hand sides of rules start with a combinator symbol $c$ where $c \in \mathcal{F}$ is followed by the types of arguments required to obtain the type on the left hand side of the rule by applying the combinator.
When (CL)S constructs a tree grammar, we have a word $M \in \mathcal{L}_{\tau}(G)$.
The computed grammar $G$ is \textit{sound} because the word $M$ is well-typed term.
Furthermore, $G$ is \textit{complete} because all requested well-typed terms are words of the grammar derived for the target type $\tau$.


\subsection{Scala Implementation}
The integration of the (CL)S algorithm into Scala allows simple specification of combinators \cite{Bessai.2018}.
A typical type specification of the repository $\Gamma$ for two combinators describing a $start$ position and an $up$ movement in a game is
\begin{align*}
  \Gamma = \{ &start : Pos(3, 4),\\
              &up : (Pos(3, 4) \to Pos(3, 3)) \cap (Pos(3, 3) \to Pos(3, 2)) \}.
\end{align*}
Here, arrows are function types and the binary intersection type operator $\cap$ means that a combinator has two types simultaneously.
Similar to dependent types \cite{brady2017type}, specifications can include arbitrary constants and types can encode precomputed function tables.
This specification mechanism is Turing complete in general \cite{DBLP:conf/csl/DudderMRU12}, but in practice we use some restrictions, rendering the existence of terms for the type inhabitation problem decidable.
In the current version, (CL)S accepts specifications in almost mathematical notation, allowing to state the example for $\Gamma$ above as:

\begin{lstlisting}[numbers=none]
val Gamma = Map("start" -> 'Pos('3, '4),
                "up" -> ('Pos('3, '4) =>: 'Pos('3, '3)) :&:
                        ('Pos('3, '3) =>: 'Pos('3, '2)))
\end{lstlisting}
It can also extract type information from combinators with implementations attached to them, allowing to enter the combinator $up$ from $\Gamma$ according to the Scala representation in Listing~\ref{scala}.
We obtain the specification with native and semantic types: $(Pos(3, 4) \to\ Pos(3, 3)) \cap\ (Pos(3, 3) \to Pos(3, 2)) \cap (Player \to Player)$.
\begin{lstlisting}[numbers=none, caption={Scala representation of a combinator with native and semantic types}, captionpos=b, label={scala} ]
@combinator object up {
  def apply(player: Player): Player = player.goUp()
  val semanticType =
    ('Pos('3, '4) =>: 'Pos('3, '3)) :&:
    ('Pos('3, '3) =>: 'Pos('3, '2)))      }
\end{lstlisting}
The intersection type operator is represented by $:\&:$ and the function types by $=>:$.
The signature of apply is automatically translated from its native Scala type.
Additional semantic type information is taken as-is and used only to impose more conditions on the use of $up$, which are user specified.
The term returned for question $\Gamma \vdash ? : Pos(3, 3)$ is $up(start)$, which, when providing combinator implementations, is automatically translated to the method calls \texttt{up.apply(start.apply)}.
The following tree grammar is the result of the inhabitation:
\begin{center}
\hspace{-2em}$\begin{aligned}
	G = \{
	& Pos(3, 4) \mapsto\  \{start()\}, \\
	& Pos(3, 3) \mapsto\  \{up(Pos(3, 4))\}, \\
	& Pos(3, 2) \mapsto\  \{up(Pos(3, 3))\}
	~\}
\end{aligned}$
\end{center}



\newcommand{\iht}{inhabTree}
\newcommand{\parameters}{args}
\newcommand{\foolist}{xorSet} 
\newcommand{\cList}{constrSet}
\newcommand{\typeFunction}{ty~}
\section{CLS-SMT}
\label{cls-smt}

This section describes the key aspects of CLS-SMT.
The production rules in the grammar are used to formulate SMT constraints by using uninterpreted functions.
Any SMT model satisfying the given constraints represents a tree, which is necessarily a word of the tree grammar.


We define a data structure that represents applicative terms and show how a (CL)S tree grammar can be translated to an adequate SMT formulation.






\begin{definition}
\label{def:tree}
{(Inhabitant Tree)}

An inhabitant tree is a binary tree over integers. Let n denote the finite number of combinators used in the tree grammar and $C \subset \mathbb{N}$ range over $\{1 , ... , n\}$. With $c \in C$, an inhabitant tree is defined as follows:
\begin{center}$\iht~=~0~(leftChild~\iht)~(rightChild~\iht)~|~c$\end{center}
\end{definition}

Accordingly, the tree's alphabet of vertex labels $\Sigma_V$ is $\{0\}\cup C$. A vertex labeled $0$ is called application node and denoted by $@$. An $@$ node has exactly two children (i.e. 0 is a binary symbol), the function is the left child and argument is the right child. All elements of $C$ are constants so that $@$ nodes are the only elements of the tree that are allowed to have children. A combinator with $n$ arguments is represented by a tree that consists of (at least\footnote{more @ nodes could be contained in the subtrees representing the arguments}) $n$ application nodes and the combinator symbol on the leftmost leaf. The $n$-th argument of a combinator is the right child of the combinators $n$-th parent.
As an example, we consider the term $((c~(arg1))~arg2)$, which represents the application of the binary combinator $c$ to the arguments $arg1$ and $arg2$.

We assume that $c$ is encoded as $1$, $arg1$ as $2$ and $arg2$ as $3$. The corresponding inhabitant tree is $0~(leftChild~(0~(leftChild~1)~(rightChild~2))~(rightChild~3)$. A visual representation is as follows:

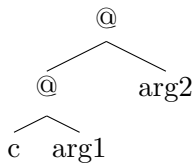
\Tree[.@ [.@ [.c ] [.arg1 ] ]  [.arg2 ] ]

\subsection{Constraint Representation}

Due to the completeness of the inhabitation algorithm, there is at least one applicative term that can be build from a non-empty tree grammar.
Thus, an SMT encoding of the tree grammar on its own will always be satisfiable.
There is no need for an encoding of the subtyping relation because subtyping is considered in the inhabitation algorithm.
Accordingly, the tree grammar only contains nonterminals representing types and there is a production rule for every nonterminal used.

Let $V$ be the finite set of vertices.
The labelling function $inhabitant: V \mapsto \Sigma_V$ can be used for a total representation of a tree if the rules given in Definition \ref{def:tree} are respected.
We use the production rules in the tree grammar to formulate structural constraints on the tree.
Let $n \in N$ and $N$ denote the set of nonterminals of the grammar.
We introduce the partial function $ty: V \mapsto N$, which maps vertices of a tree to a nonterminal representing a type.
The information provided by a production rule of the tree grammar can now be used to systematically build constraints for the corresponding subtree.
We consider the production rule \{$\alpha$ $\mapsto$ \{(c($\beta_1$, $\beta_2$)\}\} and its incomplete tree representation that is supplemented with the associated nonterminals:


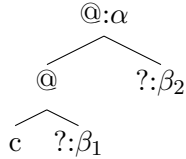
\Tree[.@:$\alpha$ [.@ [.c ] [.?:$\beta_1$ ] ]  [.?:$\beta_2$ ] ]

Its possible to derive the following constraints from this production rule. Let $i$ denote the root node of the applicative composition of the combinator and its arguments.
If node $i$ has type represented by nonterminal $\alpha$ then the vertex \texttt{(leftChild (leftChild i))} must be c, the first argument (at position \texttt{(rightChild (leftChild i))}) must be typed according to $\beta_1$ and the second argument (at \texttt{(rightChild i)}) must be typed corresponding to $\beta_2$.
The constraints for subtrees denoted by $\beta_1$ and $\beta_2$ can be formulated accordingly.
Following this approach, the contents of a tree grammar can be translated into SMT constraints.
Adequate assertions are formulated and supplied to the SMT solver to find implementations for the uninterpreted functions $inhabitant$ and $ty$.
We currently use Z3 from Microsoft Research \cite{de2008z3} to solve our formulation with the background theory LIA \cite{SMTLIBInitiative} (i.e. the linear fragment of the theory of Integers).
A more detailed look at the translation will be given in the next section.

\subsection{Grammar Translation}
We translate the grammar by applying \textsc{Translate_Production_Rule} shown in Algorithm \ref{gRuleTranslation} to every production rule of the grammar.
The algorithm produces SMT boolean expressions that must evaluate to true for all vertices of a valid tree.
We make use of the aforementioned functions $inhabitant$ and $ty$ to formulate these constraints.
The set of constraint functions is incorporated in an assertion with a $forall$ expression where the universal quantified variable $i$ represents the vertices. Consequently, every solution found by the SMT solver must be a word of the grammar.

Inside \textsc{Translate_Production_Rule}, the function \texttt{Translate_Combinator} is applied to every possible combinator listed in this specific production rule.
The resulting set of boolean expressions is joined with the \texttt{xor} connective as we must use one combinator subtree exclusively at a given type annotated vertex.
For the sake of readability, we assume that \texttt{xor} and \texttt{and} are applicable to sets.

An $n$-ary combinator is translated by using the universal quantified variable $i$ and its associated children to describe the vertices of the respective subtree.
The labelling is formulated by placing constraints on the $ty$ and $inhabitant$ functions.
We reverse the list of nonterminals $\parameters$~that describes the required types of a combinator's arguments in order to address the structure of inhabitant trees.
That way, we can start at the root node of the current subtree and build successive address terms for each loop iteration by applying \texttt{leftChild} to the current address term.
The complete structure of the subtree must satisfy all constraints that were produced in the loop, so we return the corresponding conjunction.
After translating the grammar rules, we also include a root node constraint.
It states that $ty$ must map node 1 of the tree to the nonterminal representing the synthesis goal type.

\begin{algorithm}[H]
\caption{Production Rule Translation}\label{gRuleTranslation}
 \label{alg1}
\begin{algorithmic}
\Function{Translate_Production_Rule}{$typeId, values$}
\State $\foolist \gets \varnothing$
\ForAll{$(combinator, parameters)$~in~values}
\State $cTransl \gets \textsc{Translate_Combinator}$(combinator, parameters)$ $
\State $\foolist~\gets~\foolist~\cup~cTransl$
\EndFor\label{fooloop}
\State \textbf{return} \texttt{(ite (= (\typeFunction i) $typeId$) (xor $\foolist$) true)}
\EndFunction
\Statex
\Function{Translate_Combinator}{$combinator, \parameters$}
\State $\cList \gets \varnothing$
\State $currentAddress \gets i$
\State $pList \gets \parameters.reverse$
\ForAll{p in pList}
\State $\cList \gets \cList~\cup~$\texttt{(= (\typeFunction (rightChild $currentAddress$)) p)}
\State $\cList \gets \cList~\cup~$\texttt{(= (inhabitant $currentAddress$) 0)}
\State $currentAddress \gets $ \texttt{(leftChild $currentAddress$)}
\EndFor\label{forList}
\State $combinatorConstraint \gets $ \texttt{(= (inhabitant $currentAddress$) combinator)}
\State $combinedSet \gets combinatorConstraint \cup \cList$
\State \textbf{return} \texttt{(and ($combinedSet$))}
\EndFunction
\end{algorithmic}
\end{algorithm}

Any tree model $M^*$ that satisfies these constraints represents a word $M$ of the grammar and every word $M$ can be translated to a model $M^*$ that satisfies these constraints.
The translation is straight-forward and is thus be omitted.
Let $\varphi$ denote the conjunction of the constraints and $\tau$ denote the inhabitation target type, then: $M^* \vDash_{LIA} \varphi \Leftrightarrow M \in \mathcal{L}_{\tau}(G)$.


\section{Experiments}
In this section, we discuss the advantages and the usefulness of our approach by means of a composition of sort programs and a path finding scenario.

\newcommand{\doubleType}{double}
\newcommand{\alphaType}{\alpha}
\newcommand{\rarr}{\rightarrow}

\subsection{Sort}
We consider a small repository $\Gamma$ shown in Fig. \ref{sortrepo} that can be used to compose sort programs.
It contains a sort combinator for lists that applies a function to each element before performing the sorting.
The $id$ combinator typed $\alpha \rarr \alpha$ can be used if we want to sort the unmodified list values.
Moreover, the inverse function can be applied to double values.
Further combinators could include the $abs$ function to compare absolute values or a $dist$ combinator to calculate the distance to a given value.

\begin{figure}[H]
\begin{center}
\hspace{-2em}$\begin{aligned}
    \Gamma = \{
		 \;& values : List(\doubleType), \\
    \;& id  : \alpha \rarr \alpha, \\
    \;& inv : \doubleType \rarr \doubleType, \\
    \;& sortmap : (\alphaType \rarr \alphaType) \rarr List(\alphaType) \rarr SortedList(\alphaType), \\
    \;& min : \doubleType \rarr SortedList(\doubleType) \rarr minimal \cap \doubleType, \\
    \;& default : \doubleType 
		~\}
\end{aligned}$
\end{center}
\caption{Repository for the sort example}
\label{sortrepo}
\end{figure}


In some cases, it might be required to sort a $double$ list and additionally determine its minimal value. 
The corresponding combinator $min$ will be implemented by extracting the first value of a sorted list (assuming that we always sort in an ascending order).
The result type of $min$ is an intersection of $minimal$ and $\doubleType$.
For empty lists, a default value will be returned. 
In this example, such a value is held in the component $default$, which has the type $double$.
The inhabitation request $\Gamma \vdash ? : minimal \cap \doubleType$ yields the following grammar $G$:

\begin{figure}[H]
\begin{center}
\hspace{-2em}$\begin{aligned}
	G = \{
	& SortedList(\doubleType) \mapsto\  \{sortmap(\doubleType \rarr \doubleType, List(\doubleType))\}, \\
	& minimal \cap \doubleType \mapsto \{id (minimal \cap \doubleType), min(\doubleType, SortedList(\doubleType)) \}, \\
	& double \mapsto \{id(\doubleType), default(), inv(\doubleType), min(\doubleType, SortedList(\doubleType))\}, \\
	& double \rarr double \mapsto \{id(), inv()\} \\
	& List(double) \mapsto \{id(List(\doubleType)), values()\}
	~\}
\end{aligned}$
\end{center}
\caption{Tree grammar for the sort example, $\Gamma \vdash ? : minimal \cap \doubleType$ }
\label{tGrammarSort}
\end{figure}

A double value can be formed by applying $id$ or $inv$ to any term with type double.
Obviously, terms like $inv$ and $id$ can be applied an arbitrary number of times to arguments of type double.
Thus, the range of terms with type double is infinite.
Moreover, a term typed $minimal \cap \doubleType$ can also be used as the first argument of the $min$ operator.
The grammar describes all well-formed solutions that comply to the target type.
However, it is clearly not desirable to compose infinite range of trivial solutions.
With extensions formulated as SMT constraints, we can further filter the result set without specializing $\Gamma$ too much.

In order to avoid trivial solutions, we specify $id$ and $inv$ to be used only as arguments.
Moreover, the first argument of $min$ must be a terminal.
Given the indices 2, 3 and 5 for the combinators $id$, $min$ and $inv$, the following assertions are added to the SMT script:
\begin{lstlisting}
(assert (forall ((i Int)) (not (= (inhabitant (leftChild i)) 2))))
(assert (forall ((i Int)) (not (= (inhabitant (leftChild i)) 5))))
(assert (forall ((i Int))
	(ite (= (inhabitant (leftChild i)) 3)
		(not (= (inhabitant (rightChild i)) 0)) true)))
\end{lstlisting}

With these constraints at hand, only two valid solutions are found for the inhabitation request $\Gamma \vdash ? : minimal \cap \doubleType$:\begin{center}
$((min~default)~((sortmap~inv)~values))$ and \\
$((min~default)~((sortmap~id)~values))$
\end{center}

The combinator $min$ is applied to the terms yielded by the combinators $default$ and $sortmap$.
For this particular example, the combinator mapping in the table shown below was used.
In order to illustrate the first result term as a tree, we use the following labelling pattern: \\
$combinator~name~:~(vertex~id,~combinator~id)$ \\

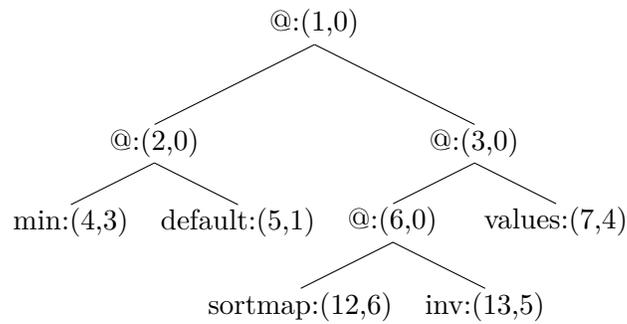
\begin{minipage}{0.35\textwidth}
\begin{center}
\begin{tabular}{|l|c|r|}
\hline
 name & id \\
\hline
default & 1 \\
id & 2 \\
min & 3 \\
values & 4 \\
inv & 5 \\
sortmap & 6	\\
\hline
\end{tabular}
\end{center}
\end{minipage}
\begin{minipage}{0.5\textwidth}
\Tree[.@:{(1,0)} [.@:{(2,0)} [.min:{(4,3)} ] [.default:{(5,1)} ] ] !\qsetw{1cm} [.@:{(3,0)} [.@:{(6,0)} [.sortmap:{(12,6)} ] [.inv:{(13,5)} ] ] !\qsetw{2cm} [.values:{(7,4}) ] ] ]
\end{minipage}



\subsection{Labyrinth Example}

In the following labyrinth example, it is possible to go $up$, $down$, $left$ or $right$, if the new position is not occupied by obstacles \cite{Bessai.2018}.
Fig.~\ref{fig:lab} illustrates a 3 $\times$ 4 labyrinth example.
The starting position is $(0, 2)$ (shown as $\bullet$) and the goal position $(1, 0)$ (shown as $\bigstar$).

\begin{figure}[H]
\begin{center}
\begin{tabular}{r|C{0.4cm}|C{0.4cm}|C{0.4cm}|}
  & 0                 & 1                 & 2 \\\hline
0 & \cellcolor{black} & $\bigstar$        & \cellcolor{black} \\\hline
1 &                   &                   &  \\\hline
2 &	$\bullet$         &	\cellcolor{black} &\\\hline
3 &                   &                   &\\\hline
\end{tabular}

\caption{Labyrinth example}
\label{fig:lab}
\end{center}

\end{figure}
The repository with typed combinators for this example is represented in Fig.~\ref{repo}.

\begin{figure}[H]
\begin{center}
\hspace{-2em}$\begin{aligned}
    \Gamma_{Lab} = \{
           ~left  :\;& (Pos(1, 1) \to Pos(0, 1)) \cap Pos(2, 1) \to Pos(1, 1))\ \cap \\
                    & (Pos(1, 3) \to Pos(0,3)) \cap (Pos(2, 3) \to Pos(1, 3)),\\
           right :\;& (Pos(0, 1) \to Pos(1, 1)) \cap (Pos(1, 1) \to Pos(2, 1))\ \cap \\
                    &(Pos(0, 3) \to Pos(1,3)) \cap (Pos(1, 3) \to Pos(2, 3)),\\
            up :\;& (Pos(0, 3) \to Pos(0, 2)) \cap (Pos(2, 3) \to Pos(2, 2))\ \cap \\
            	&(Pos(1, 1) \to Pos(1, 0)) \cap (Pos(0, 2) \to Pos(0, 1))\ \cap \\
            	&(Pos(2, 2) \to Pos(2, 1)), \\
           down  :\;& (Pos(1, 0) \to Pos(1, 1)) \cap (Pos(0, 1) \to Pos(0,2))\ \cap\\
           		&(Pos(2, 1) \to Pos(2, 2)) \cap (Pos(0, 2) \to Pos(0,3))\ \cap \\
           		&(Pos(2, 2) \to Pos(2, 3)),\;\; \\
           start :\;& Pos(0, 2)
           ~\}
\end{aligned}$
\end{center}
\caption{Repository for the labyrinth example shown in Fig.~\ref{fig:lab} }
\label{repo}
\end{figure}
The combinators \textit{up, down, left}, and \textit{right} can be used to go from position $Pos(x,y)$ to an accessible neighbouring position.
The types $Pos(x,y)$ represent the column and row positions.
For example, combinator \textit{left} can be used to go from position $Pos(1,1)$ to position $Pos(0,1)$ 
as well as from $Pos(2,1)$ to $Pos(1,1)$, from $Pos(1,3)$ to $Pos(0,3)$, and from $Pos(2,3)$ to $Pos(1,3)$.
The combinator \textit{start} provides the starting position.

To get all possible paths from start (0,2) to goal position (1,0), we ask for:
$$\Gamma \vdash ?: Pos(1,0)$$
For this goal position the algorithm computes the grammar shown in Fig.~\ref{tGrammar}.
\begin{figure}[H]
\begin{center}
\hspace{-2em}$\begin{aligned}
	G = \{
	& Pos(1,0) \mapsto\  \{up(Pos(1,1))\}, \\
	& Pos(1,1) \mapsto\  \{right(Pos(0, 1)),
	\ left(Pos(2,1)),
	\ down(Pos(1,0))\}, \\
	& Pos(1, 1) \mapsto\  \{up(Pos(0, 2)),
	\ left(Pos(1, 1))\},\\
	&Pos(2, 1) \mapsto\  \{up(Pos(2, 2)),
	\ right(Pos(1, 1))\},\\
	&Pos(2, 2) \mapsto\  \{down(Pos(2, 1)),
	\ up(Pos(2, 3))\},\\
	&Pos(0, 1) \mapsto\  \{up(Pos(0, 2)),
	\ left(Pos(1, 1))\},\\
	&Pos(0,3) \mapsto\ \{ down(Pos(0, 2)),
	\ left(Pos(1, 3))\},\\
	&Pos(0, 2) \mapsto\  \{down(Pos(0, 1)),
	\ up(Pos(0, 3)),
	\ start()\},\\
	&Pos(1, 3) \mapsto\  \{left(Pos(2,3)),
	\ right(Pos(0, 3))\},\\
	&Pos(2, 3) \mapsto\  \{down(Pos(2, 2)),
	\ right(Pos(1, 3))\}
	\}
\end{aligned}$
\end{center}
\caption{Tree grammar for the labyrinth example}
\label{tGrammar}
\end{figure}

For the path going  $up$, $right$, and $up$ the algorithm constructs a term $up(right(up(start)))$.
In this example, there are also terms that represent trivial paths with cycles.
For example:
\begin{center}
$\begin{aligned}
\;&up(right(up(down(up(down(up(start))))))),\\
\;&down(up(up(right(up(start))))),...
\end{aligned}$
\end{center}
By means of SMT solvers, we can restrict the number of solutions computed by (CL)S in order to avoid trivial terms.
For example, we can decide, which combinators have to be used and how often.
As presented in Section~\ref{cls-smt} we translate the computed tree grammar (s. Fig.~\ref{tGrammar}) to SMT expressions by means of algorithm~\ref{alg1}.

In order to filter the inhabitants, we consider domain-specific constraints.
We are able to select, which combinators should be used in the solution.
For instance, Fig.~\ref{assert} shows a formula that states a term should not include combinator $down$ (translated as \texttt{(= (inhabitant i) 1)}).
This way, we constrain the usage of certain combinator.
In this particular example (see Fig.~\ref{fig:lab}), we might want to avoid the $down$ combinator, because the robot has to get to the top-right goal position.

\begin{figure}[H]
\begin{lstlisting}
(assert (forall ((i Int)) (not (= (inhabitant i) 1))))
\end{lstlisting}
\caption{Assertion for filtering of combinator}
\label{assert}
\end{figure}
We reduce the number of cycles and define the order of usage of the combinators in order to avoid unnecessary paths.
For example, we can formulate a constraint that forbids the application of combinator $down$ (index 1) to combinator $up$ (index 2) and vice versa.
The same applies to combinators $left$ (index 3) and $right$ (index 4).
Fig.~\ref{order} shows the definition of this rule.

\begin{figure}[H]
\begin{lstlisting}
(assert (forall ((i Int))
 (and
  (not (and (= (inhabitant (leftChild i)) 3)
  (= (inhabitant (leftChild (rightChild i))) 4)))
  (not (and (= (inhabitant (leftChild i)) 4)
  (= (inhabitant (leftChild (rightChild i))) 3)))
  (not (and (= (inhabitant (leftChild i)) 2)
  (= (inhabitant (leftChild (rightChild i))) 1)))
  (not (and (= (inhabitant (leftChild i)) 1)
  (= (inhabitant (leftChild (rightChild i))) 2))))
 ))
\end{lstlisting}
\caption{Formula for definition of order}
\label{order}
\end{figure}

\section{Related Work}
\subsubsection*{Type-theoretical specification}
There are various approaches to solve synthesis problems by means of type theory.
For instance, Polikarpova et al. synthesized recursive functions satisfying a specification in the form of polymorphic refinement types \cite{PKS16}.
Zdancewic et al. demonstrated that examples in example-directed synthesis can be interpreted as refinement types \cite{Zdancewic16}.
They provided an example-based specification language by using intersection types with singletons.
In contrast, (CL)S expresses semantic specifications with intersection types.
Kuncak et al. used type inhabitation in the simply typed lambda calculus to support developers by generating a list of valid expressions of a given type for code completion \cite{GKKP13}.

\subsubsection*{SMT}
In the last decades, there have been many approaches using SMT solvers for synthesis.
A common property of those methodologies is the use of syntactic constraints and a correctness specification.
In 2006, preliminary work in template-based synthesis was undertaken by Solar-Lezama et al. \cite{SolarLezama06}.
In \emph{Sketching}, a partial implementation is given and synthesis completes missing parts by considering a specification of the desired functionality \cite{SolarLezama06}.
Following this idea, loop-free bitvector programs \cite{Gulwani11} and deobfuscating programs \cite{JGST10} were synthesized in a component-based manner.
In contrast to our work, desired functionality and components were specified as logical relations between the input and output variables \cite{Gulwani11,JGST10}.
Another approach in SMT based synthesis is programming by examples.
A user specifies the behaviour of the desired program by a number of input-output examples \cite{GPS17}.
Singh and Gulwani transformed strings and data types in spreadsheets \cite{GHS12,SinghGulwani16} and Udupa et al. were able to synthesize protocols from a given skeleton and examples \cite{Udupa13}.

In 2013, a number of researchers picked up the main ideas of the projects above to formulate the problem of syntax-guided synthesis (SyGuS) \cite{SyGuS13}.
The Counterexample-Guided Inductive Synthesis (CEGIS) architecture describes how SyGuS problems can be tackled by learning from counterexamples provided by a verification oracle, which is often implemented by off-the-shelf SMT solvers \cite{SyGuS13}.

Most of the synthesis algorithms based on CEGIS variants are solving $\exists\forall$-formulas iteratively using SMT solvers \cite{SyGuS13}.
Similar to Reynolds et al. we consider synthesis as a theorem-proving problem.
In our case, the problem is solved in combinatory logic and later refined by a SMT solver, whereas in \cite{Reynolds17} the problem is solely solved within the SMT solver.
The main difference is the way of specification.
Like in many traditional synthesis approaches \cite{Zdancewic16,GHS12,SinghGulwani16,Udupa13}, targets in \cite{Reynolds17} are specified by using properties of executed programs.
More specifically, relations on inputs and outputs are defined.
This allows for a fine-granular specification on program behaviour, but it is hard to control the structure of synthesized programs.
It can also be hard to specify the program behaviour in the SMT solver, which becomes especially apparent in the presence of side effects or exceptions.
In (CL)S, these concerns are hidden behind the interfaces of types.
Types are particularly easy to define, because they already exist in most programming languages and do not need to be specified just for synthesis.
They can encode taxonomic concepts via semantic types and subtyping, which is usually a very natural way of expression \cite{Steffen:97}.
In future work, it might be interesting to consider bridging the gap between behavioural and type-based specifications.
In particular, the approach in \cite{Reynolds17} could be used to synthesize the implementation for individual combinators, which are then composed by (CL)S and CLS-SMT.
\section{Conclusion}
In our work we combined Combinatory Logic Synthesis and Satisfiability Modulo Theories in a tool called CLS-SMT.
In this way, we are able to compensate limitations of one technology by taking advantage of the other and vice versa.
The synthesis framework (CL)S generates a tree grammar from a given repository of typed components that contains domain-specific knowledge.
We should emphasize that the tree grammar is \emph{complete} and describes all well-formed solutions.
CLS-SMT translates the grammar into SMT formulas and further domain-specific constraints are added.
The SMT solver Z3 finds a model considering the translated tree grammar and constraints.

By having further constraints formulated as SMT formulas, we are able to restrict inhabitants without restricting the types of the (CL)S component repository.
That way, we benefit from the expressiveness of first-order logic and background theories.
Combinatory logic synthesis reduces the search space of the SMT solver.
In general, SMT considers this structure of the programs, whereas components in (CL)S contain domain-specific details.
Combinatory logic with intersection types is a Turing complete formalism that allows to define semantic taxonomies based on subtyping \cite{DBLP:conf/isola/BessaiDDMR16}.


Although SMT solvers are highly efficient through decades of research and improvements, handling quantified formulas is still challenging.
Congruence Closure with Free Variables (CCFV) \cite{Barbosa17} is a framework that is based on the E-ground (dis)unification problem and unifies major instantiation techniques in SMT solving.
Experimental evaluation shows that CCFV improved the performance of the solvers CVC4 and veriT significantly, so that the former outranks the state-of-the-art in instantiation based SMT solving.
Within our research, the replacement of solvers is possible with reasonable effort due to the SMT-LIB standard.
Further performance enhancements could be achieved by exploring the usage of data types to express the tree.

We have applied CLS-SMT to synthesize sort programs and motion plans.
Motion planning problems are an interesting topic for program synthesis because of the associated scaling problems.
Our examination shows that synthesis of small motion plans is successful.
On the other hand, we found that larger examples do not scale properly.
Our approach is well-suited for motion plans with up to $10 \times 10$ tiles. 
Scaling problems do not apply to other use cases such as the sort example.
Future work considers an investigation of motion planning problems with multiple robot instances and obstacles.

\medskip

\textbf{Acknowledgement.} The work presented in this paper was partly funded by the GRK 2193 (\url{www.grk2193.tu-dortmund.de/de/}) and the Center of Excellence for Logistics and IT (\url{www.leistungszentrum-logistik-it.de/}) located in Dortmund.

\bibliographystyle{eptcs}
\bibliography{bibliography}
\end{document}